\documentclass[]{aastex631}
\usepackage[utf8]{inputenc}
\usepackage{amsmath}
\usepackage{float}
\usepackage{CJK}
\usepackage{graphicx}
\usepackage{multirow}
\usepackage{comment}

\begin{document}

\begin{CJK*}{UTF8}{gbsn}  

\title{Simultaneous Proton and Electron Energization during Macroscale Magnetic Reconnection}

\author[0009-0009-7643-6103]{Zhiyu Yin (尹志宇)}
\affiliation{Department of Physics, University of Maryland, College Park, MD 20742, USA}
\affiliation{IREAP, University of Maryland, College Park, MD 20742, USA}

\author[0000-0002-9150-1841]{J. F. Drake}
\affiliation{Department of Physics, University of Maryland, College Park, MD 20742, USA}
\affiliation{IREAP, University of Maryland, College Park, MD 20742, USA}
\affiliation{Institute for Physical Science and Technology, University of Maryland, College Park, MD 20742, USA}
\affiliation{Joint Space Science Institute, University of Maryland, College Park, MD 20742, USA}

\author[0000-0002-5435-3544]{M. Swisdak}
\affiliation{IREAP, University of Maryland, College Park, MD 20742, USA}
\affiliation{Joint Space Science Institute, University of Maryland, College Park, MD 20742, USA}

\begin{abstract}

The results of simulations of magnetic reconnection accompanied by electron and proton heating and energization in a macroscale system are presented. Both species form extended powerlaw distributions that extend nearly three decades in energy. The primary drive mechanism for the production of these nonthermal particles is Fermi reflection within evolving and coalescing magnetic flux ropes. While the powerlaw indices of the two species are comparable, the protons overall gain more energy than electrons and their power law extends to higher energy. The power laws roll into a hot thermal distribution at low energy with the transition energy occurring at lower energy for electrons compared with protons. A strong guide field diminishes the production of non-thermal particles by reducing the Fermi drive mechanism. In solar flares, proton power laws should extend down to 10's of keV, far below the energies that can be directly probed via gamma-ray emission. Thus, protons should carry much more of the released magnetic energy than expected from direct observations.

\end{abstract}


\section{Introduction} \label{sec:intro}
Magnetic reconnection drives the fast release of magnetic energy in solar flares \citep{lin71,Emslie04,Emslie05,Emslie12}, in Earth's space environment \citep{Oieroset02,Ergun20} and in the solar wind \citep{Desai22,Phan22}. These intense events transform magnetic energy into particle energy, giving rise to a substantial population of non-thermal particles that display a power-law tail in their distribution functions. 

Instrumental data from the Reuven Ramaty High Energy Solar Spectroscopic Imager (RHESSI) and the Atmospheric Imaging Assembly on the Solar Dynamics Observatory suggest that non-thermal electrons gain a considerable portion of the overall electron density in solar flares \citep{Lin03,Warmuth16}.  Measurements of non-thermal ions in such events are challenging \citep{Emslie12}.  Gamma-ray lines reveal information about energetic protons above around an MeV.  However, gamma-ray flares are uncommon and the spectra of energetic protons below an MeV are not directly measureable. The spectra of energetic ions from impulsive flares are measured in the solar wind and reveal extended powerlaws \citep{Mason07}. The proton spectra and the total energy content of these and other ion species are, however, not well characterized. On the other hand, observations of reconnection in the Earth's magnetotail \citep{Ergun20} and at the heliospheric current sheet (HCS) close to the sun reveal that protons and other ion species carry substantial energy \citep{Desai22,Phan22}. Because of the uncertainty about the energy content of protons in flares, it is imperative to develop a self-consistent simulation model that can probe magnetic reconnection in macro-scale systems relevant to solar flares and other space environments and make predictions for the energy spectra and total energy content of protons. 

The dominant drive mechanism of particle energy gain during magnetic reconnection is Fermi reflection in growing and merging magnetic flux ropes \citep{Drake06,Oka10,Dahlin14,Guo14,Li19,zhang21}. The mechanism produces power-law tails in both electron \citep{Arnold21} and ion \citep{zhang21} distributions.  Magnetic reconnection produces bent magnetic field lines that form an Alfv\'enic exhaust, carry energy away from the x-line, and transfer energy from the field to the surrounding plasma \citep{Parker57,Lin93}. Particles gain energy through repeated reflections in contracting magnetic field lines, thereby leading to the observed power-law tails in particle energy distributions \citep{Drake06, Dahlin14, Li19}. 
The rate of energy gain from Fermi reflection is proportional to a particle's energy and is therefore greatest for the most energetic particles. For this reason the most energetic particles gain the most energy, which facilitates the formation of extended powerlaw tails \citep{Arnold21}. 

We present the results from the first {\it kglobal} \citep{yin24} simulations of simultaneous electron and proton energization during magnetic reconnection in a 2D macroscale system.  The model includes the self-consistent feedback of energetic particles on the dynamics while maintaining global energy conservation. Our simulations retain Fermi reflection as the primary driver for the acceleration of both protons and electrons. Similar to the previous model by \cite{Arnold21}, \textit{kglobal} excludes kinetic-scale parallel electric fields, which frequently form in the boundary layers of magnetic reconnection but are unimportant in non-thermal particle energization \citep{Dahlin16,Dahlin17,Li19}. The model does include the macroscale electric field that boosts electron heating on entry into reconnection exhausts \citep{Egedal08,Haggerty15}. The elimination of kinetic scales, which traditionally constrain particle-in-cell (PIC) modeling, facilitates the exploration of particle energization in macroscale systems.

The implications of the \textit{kglobal} simulations are noteworthy.  The previous version of \textit{kglobal} \citep{Drake19,Arnold19} included only particle electrons and produced power-law energy spectra of energetic electrons spanning nearly three decades \citep{Arnold21}, The current version also includes particle protons and produces non-thermal spectra for both species. The proton distribution features a non-thermal tail that extends over more than three decades in energy. A derivation of the underlying equations and tests of the model can be found in \cite{yin24}.  Section \ref{sec:modelsetup} of this paper discusses the parameters of the simulations, Section \ref{sec:results} describes the results, and in Section \ref{sec:conclusion} we present our conclusions and discuss the implications for understanding reconnection-driven particle energization across the heliosphere.


\section{Simulation Model Setup}
\label{sec:modelsetup}
The simulations include four distinct plasma species: fluid protons and fluid electrons (which collectively form the MHD backbone), and particle protons and electrons which are  are represented by macro-particles that move through the MHD grid.  The particles are treated in the guiding-center limit, thus eliminating the need to resolve their respective Larmor radii. The simulations were performed within a two-dimensional (2D) spatial domain. The fluid motions are in three space direction. Particles move across the magnetic field with their $\mathbf{E}\times\mathbf{B}$ drift and along the local magnetic field at their parallel velocity. 

As with the previous version of \textit{kglobal} \citep{Arnold21}, the reconnecting component of the upstream magnetic field \( B_0 \) (along the $x$ direction) and the total proton density (the combined number density of particle and fluid protons) \( n_{i0} \) serve as normalization parameters by defining the Alfv\'en speed \( C_{A0} = B_0 / \sqrt{4\pi m_i n_{i0}} \). Because kinetic scales are excluded in the model, lengths are normalized to an arbitrary macroscale \( L_0 \) and time scales are normalized to \( \tau_A = L_0 / C_{A0} \).  Both temperatures and particle energies are normalized to \( m_i C_{A0}^2 \).  The perpendicular electric field in the simulations follows the MHD scaling \( C_{A0} B_0 / c \), while the parallel field scales as \( m_i C_{A0}^2 / (e L_0) \). Although the parallel electric field is small compared to the perpendicular component, the energy associated with the parallel potential drop over the scale \( L_0 \) is of the order of \( m_i C_{A0}^2 \), making it comparable to the available magnetic energy per particle.

The proton-to-electron mass ratio is set to $25$, but our results are insensitive to this parameter. 
The speed of light in our simulations is 30 times the Alfv\'en speed.  All species (both fluid and particle) begin with an isotropic  temperature of $0.0625$\( m_i C_{A0}^2 \), with the particles having a Maxwellian velocity distribution. The simulations are initialized with constant densities and pressures in a force-free current sheet with periodic boundary conditions.  Thus, $
B = B_0 \tanh (y/w)\hat{x} + \sqrt{B_0^2 \operatorname{sech}^2(y/w) + B_g^2} \hat{z}$ where $B_g$ is the asymptotic out-of-plane magnetic field (the guide field) and $w$, the width of the current sheet, is set to 0.005$L_0$. The initial total density of electrons ($n_e$) and protons ($n_i$) is normalized to unity, with  particles comprising $25\%$ of the density ($n_{ep}$ for particle electrons and $n_{ip}$ for particle protons) and the remaining $75\%$ in the fluid component ($n_{ef}$ for fluid electrons and $n_{if}$ for fluid protons). We demonstrate that results of the simulations are insensitive to this fraction. The simulations are conducted on grids with varying resolutions, ranging from $1024 \times 512$ to $8192 \times 4096$ grid points as shown in Table \ref{tab:initial_setting}. The number of particles is set to 100 particles per grid. The time step $\delta t$ is adjusted as the grid size is varied to
ensure that particle stepping is accurate ({\it e.g.}, $c\delta t \simeq \delta$ with $\delta$ the grid scale). The values of the time-step are given in Table \ref{tab:initial_setting}. 
 
 
%
%
%
%
%
%
%


\begin{deluxetable*}{ccccc}
\tablewidth{0pt}
\tablecaption{Parameters for Simulation Domains\label{tab:initial_setting}}
\tablehead{
  \colhead{Grid Points} & 
  \colhead{$1024 \times 512$} &
  \colhead{$2048 \times 1024$} & 
  \colhead{$4096 \times 2048$} & 
  \colhead{$8192 \times 4096$}
}
\startdata
Time Step (in $\tau_A$) & $2 \times 10^{-4}$ & $1 \times 10^{-4}$ & $5 \times 10^{-5}$ & $2.5 \times 10^{-5}$ \\
Proton Number Density Diffusion ($D_n$) & $10.5 \times 10^{-5}$ & $5.25 \times 10^{-5}$ & $2.625 \times 10^{-5}$ & $1.3125 \times 10^{-5}$ \\
Proton Pressure Diffusion ($D_p$) & $10.5 \times 10^{-4}$ & $5.25 \times 10^{-4}$ & $2.625 \times 10^{-4}$ & $1.3125 \times 10^{-4}$ \\
Hyperviscosity ($\nu_B$, $\nu_{nv}$, $\nu_n$, and $\nu_p$) & $84 \times 10^{-9}$ & $10.5 \times 10^{-9}$ & $1.312 \times 10^{-9}$ & $0.1640 \times 10^{-9}$ \\
Effective Lundquist Number ($S_\nu$) & $1.2 \times 10^{7}$ & $9.5 \times 10^{7}$ & $7.6 \times 10^{8}$ & $6.1 \times 10^{9}$ \\
\enddata
\end{deluxetable*}

In our simulations, diffusion and hyperviscosity terms are included to ensure numerical stability while reducing any high-frequency noise arising at the grid scale.  The diffusion coefficients, denoted as \(D_n\) for number density diffusion and \(D_p\) for pressure diffusion, are set to the values listed in Table \ref{tab:initial_setting}. 
A hyperviscosity  \(\nu\) is included in the magnetic field evolution equation to facilitate reconnection while minimizing dissipation at large scales. It is applied as a fourth-order Laplacian term (\(\nabla^4\)) in the equation governing the evolution of the magnetic field. The same viscosity is included in the evolution equations for the fluid proton flux, fluid proton number density and fluid proton pressure. The effective Lundquist number \(S_\nu = C_A L_0^3 / \nu\) associated with the hyperviscosity is varied to change the effective system size (the ratio of the macro to the dissipation scale).
The hyperviscosity coefficients, denoted as \(\nu_B\), \(\nu_{nv}\), \(\nu_n\), and \(\nu_p\), as well as $S_\nu$, are set to the values given in Table \ref{tab:initial_setting}. 

\section{Simulation Results} \label{sec:results}
\begin{figure}[H]
\centering
\includegraphics[width=\columnwidth]{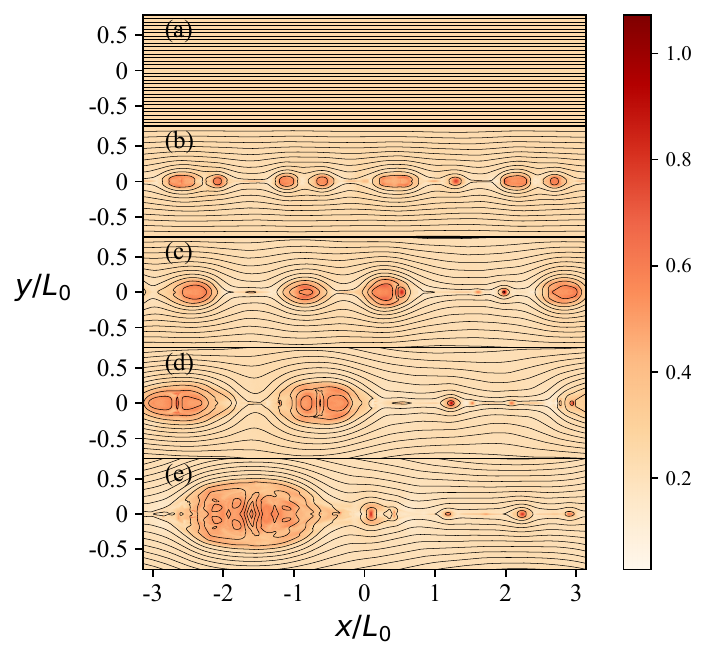}
\caption{The temporal evolution of the particle proton density in the \( x\text{--}y \) plane for \( B_g/B_0 = 0.25 \). Five times are displayed: \( t/\tau_A = 0, 4, 7, 10, \) and \( 13 \) in panels (a) to (e). In each panel, magnetic field lines are shown in black.
\label{fig:evo}}
\end{figure}

In our simulations the hyperviscosity triggers reconnection, leading to the formation of small-scale flux ropes. These flux ropes coalesce over time and eventually form system-scale structures.  Figure \ref{fig:evo} shows  the results of a simulation with \( B_g/B_0 = 0.25 \). Shown is the density of particle protons in the \( x\text{--}y \) plane at five times, \( t/\tau_A = 0, 4, 7, 10, \) and \( 13 \), in panels (a)--(e).
At $t=0$ the magnetic field reverses across a uniform current sheet as shown in panel (a) of Figure \ref{fig:evo}. As the system evolves, reconnection begins at multiple sites. The resulting reconnected magnetic field lines convect away from the x-points. The contraction of magnetic islands along the current sheet leads to the energization of particle electrons and protons trapped within them. As the simulation evolves, these islands grow and undergo coalescence, culminating in the formation of a single, large magnetic island late in time. As can be seen in Figure \ref{fig:evo} the particle proton density remains relatively uniform along the field lines due to the protons' mobility parallel to the magnetic field.

\begin{figure}[H]
\centering
\includegraphics[width=\columnwidth]{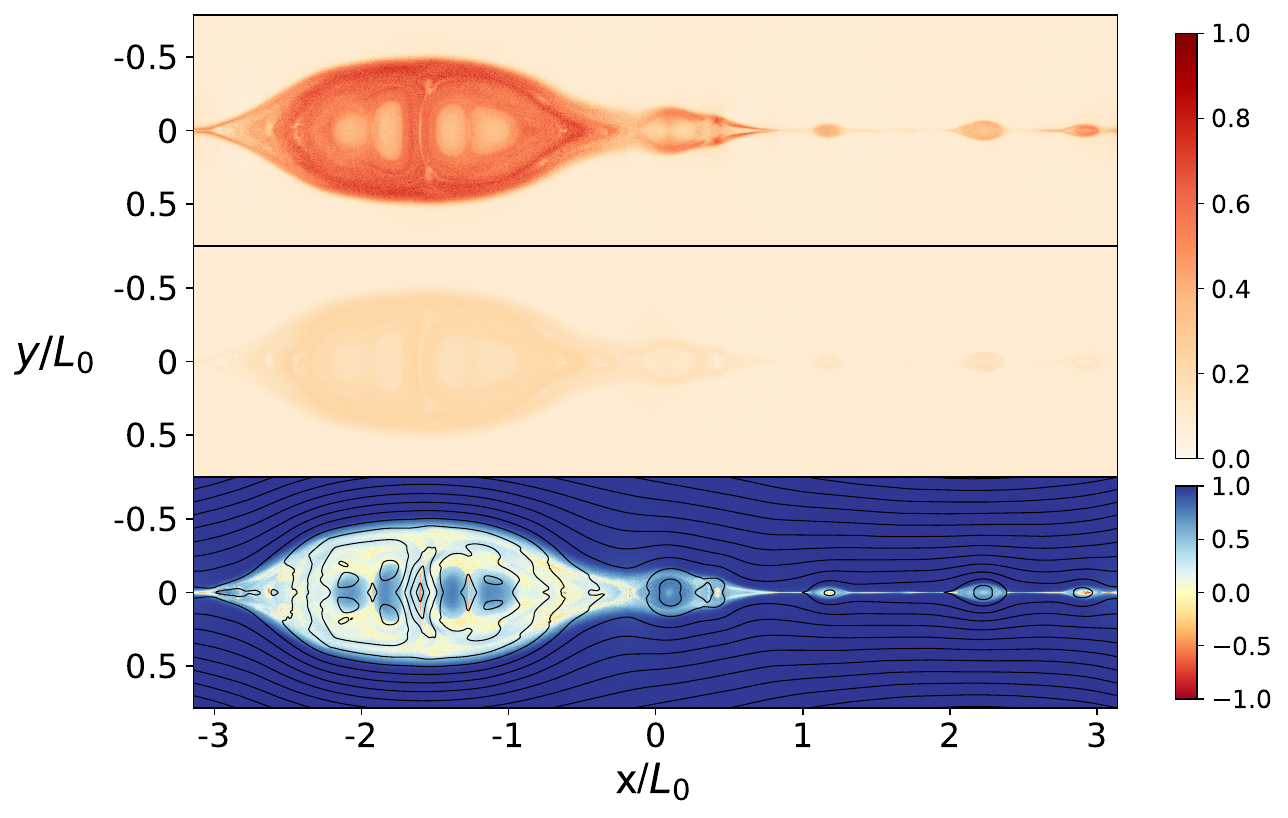}
\caption{The energy per particle of the particle protons (top) and particle electrons (middle), $\langle W \rangle$, is shown in the top two panels, respectively, for the simulation with \( B_g/B_0 = 0.25 \). The darker the color, the higher the average energy of the particles so the protons contain significantly more energy than the electrons. The firehose parameter $\alpha$ at $t = 13\tau_A$ is shown in the bottom panel. The color white corresponds to marginal stability and reveals that within flux ropes where particles have been energized, the particle feedback has significant impact on the MHD environment.}
\label{fig:fire}
\end{figure}

Figure \ref{fig:fire} shows the energy per particle (energy density divided by number density) of the particle protons and particle electrons, $\langle W \rangle$, and the firehose parameter, $\alpha$ = \(1 - 4\pi(P_{\parallel} - P_{\perp})/B^2\), at late time, \( t/\tau_A = 13 \),  from the same simulation. In the top two panels darker colors represent higher average energies per particle. Thus, the average energy of the protons is significantly greater than that of the electrons. 

The firehose parameter in the bottom panel of Figure \ref{fig:fire} is a measure of the impact of pressure anisotropy on the local magnetic tension, given by $\alpha\mathbf{B}\cdot\mathbf{\nabla}\mathbf{B}/4\pi$.  Specifically, when $\alpha$ approaches zero the magnetic tension, which is the driver of magnetic reconnection also goes to zero. Thus, the dominant feedback mechanism of energetic particles on reconnection dynamics is through the reduction of magnetic tension as measured by the firehose parameter.  In the context of the simulation from \textit{kglobal}, both \(P_{\parallel}\) and \(P_{\perp}\) are derived from the particle species since both the fluid proton and electrons are taken to be isotropic and don't impact the magnetic tension force. 
Thus, in Figure \ref{fig:fire} extensive areas within the magnetic islands approach marginal stability ($\alpha=0$), punctuated by small unstable regions (red shades). Consequently, the local magnetic tension, a crucial factor in driving reconnection and particle energy gain, is strongly reduced within the large flux rope and somewhat reduced in the smaller flux ropes. This highlights the critical importance of particle feedback on the MHD fluid dynamics.  Traditional MHD models that include test-particle dynamics do not include this feedback and thereby risk an uncontrolled increase of the particle energy.

\begin{figure}[H]
\centering
\includegraphics[width=0.8\columnwidth]{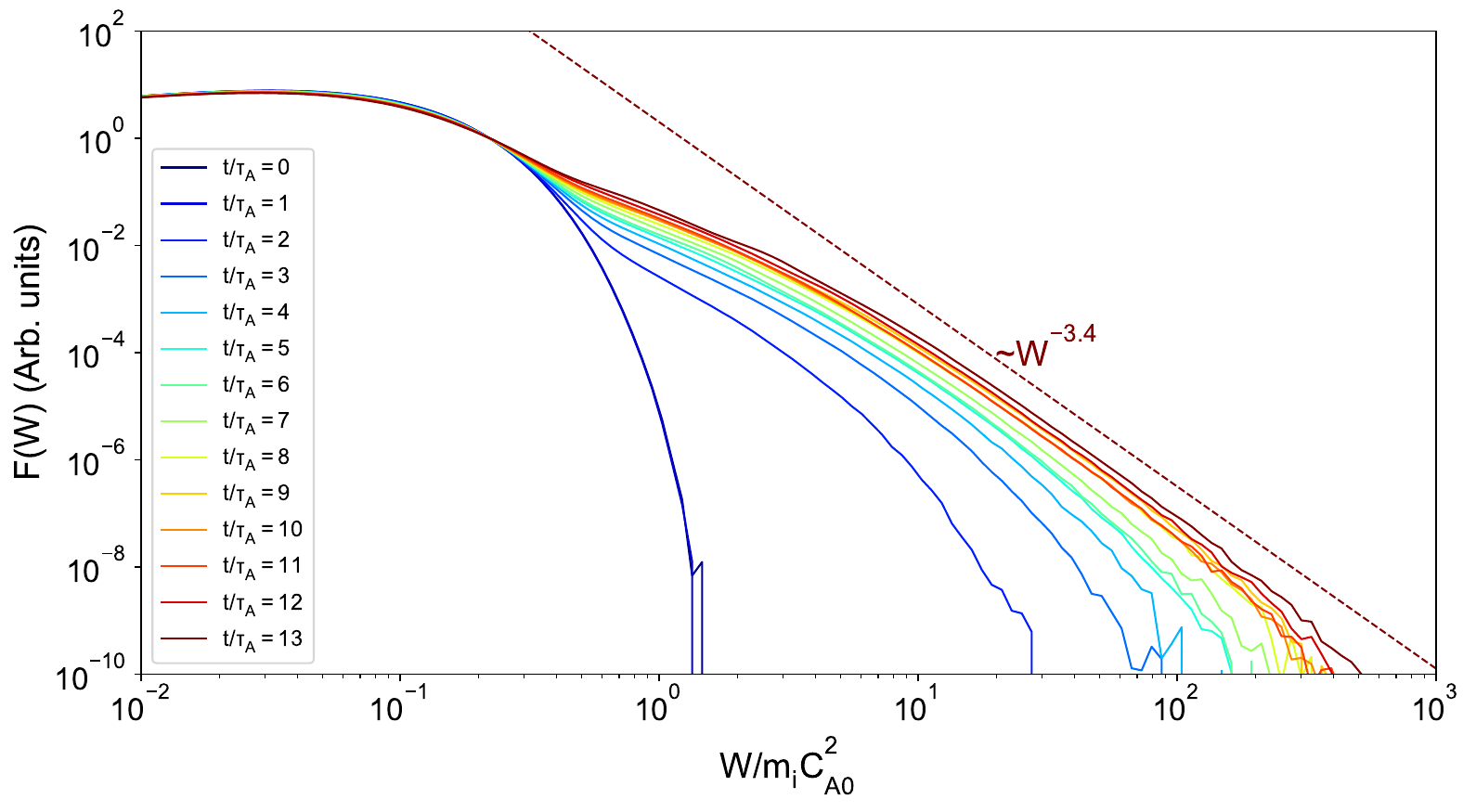}
\\
\includegraphics[width=0.8\columnwidth]{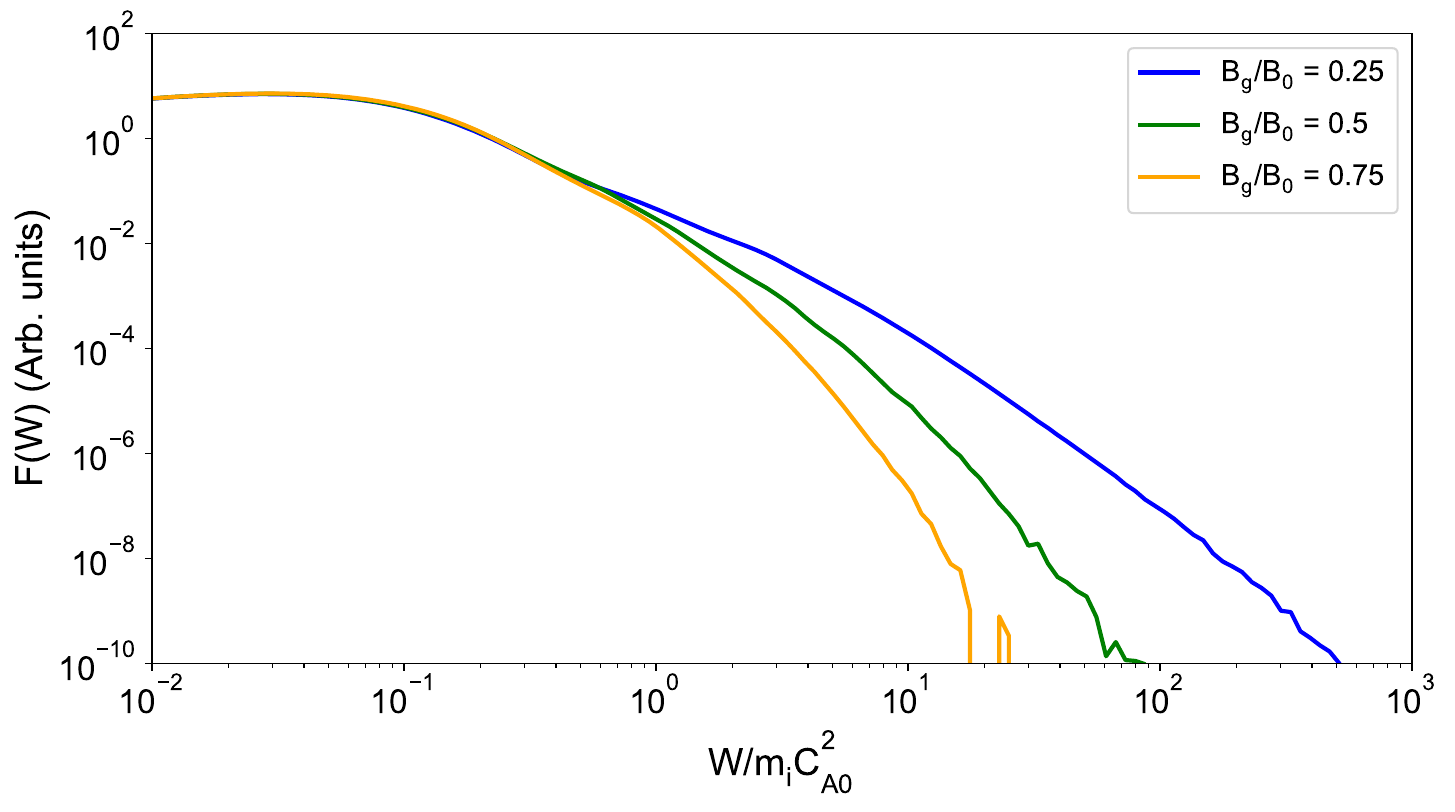}
\caption{A log-log plot of the differential number density versus the normalized energy.  Panel a shows the temporal evolution of the proton distribution in a simulation with  \( B_g / B_0 = 0.25 \). The initial Maxwellian spectrum develops an energetic tail that hardens over time and eventually stabilizes into a power law. Panel b shows the late time spectra for different guide fields. Smaller guide fields result in harder spectra that extend to higher energies.}
\label{fig:evo_bg}
\end{figure}

At late time, after a single large magnetic island dominates the current sheet, we calculate the proton energy spectra by aggregating particle counts throughout the entire simulation domain. Figure \ref{fig:evo_bg}(a) displays the differential number densities \( F(W) = dN(W) / dW \) plotted against normalized energy \( W/m_i C_{A0}^2 \) at various times for \( B_g / B_0 = 0.25 \). \( F(W) \) begins as a Maxwellian but develops a power-law tail, characterized by an index \( \delta^{\prime} \), that first appears for low energies before eventually extending to near the maximum energy observed in the domain. The powerlaw index at late time approaches 3.4. Similar behavior was observed for electrons in the earlier electron-only version of {\it kglobal} \citep{Arnold21}. Note that the particle distribution at low energy in Figure \ref{fig:evo_bg}(a) does not change significantly in time. This is not because the protons are not heated during reconnection but because the distribution in the figure is from the entire computational domain, which includes large regions of plasma upstream of the reconnecting current layer that has not been accelerated. The heating of the plasma will be more evident when we focus on the particle distributions within the large islands that develop at late time in the simulations. 

In Figure \ref{fig:evo_bg}(b) we show the late-time \( F(W) \) spectra for protons for different \( B_g \) values, measured at times when approximately equal amounts of magnetic flux have undergone reconnection. A similar figure from the electron-only version of {\it kglobal} was presented in \cite{Arnold21}. In both cases as \( B_g \) decreases, the power-law index \( \delta^{\prime} \) also decreases. This results in a harder energy spectrum and leads to the generation of a larger number of high-energy particles 
for small values of $B_g$. 

\begin{figure}[H]
\centering
\includegraphics[width=0.8\columnwidth]{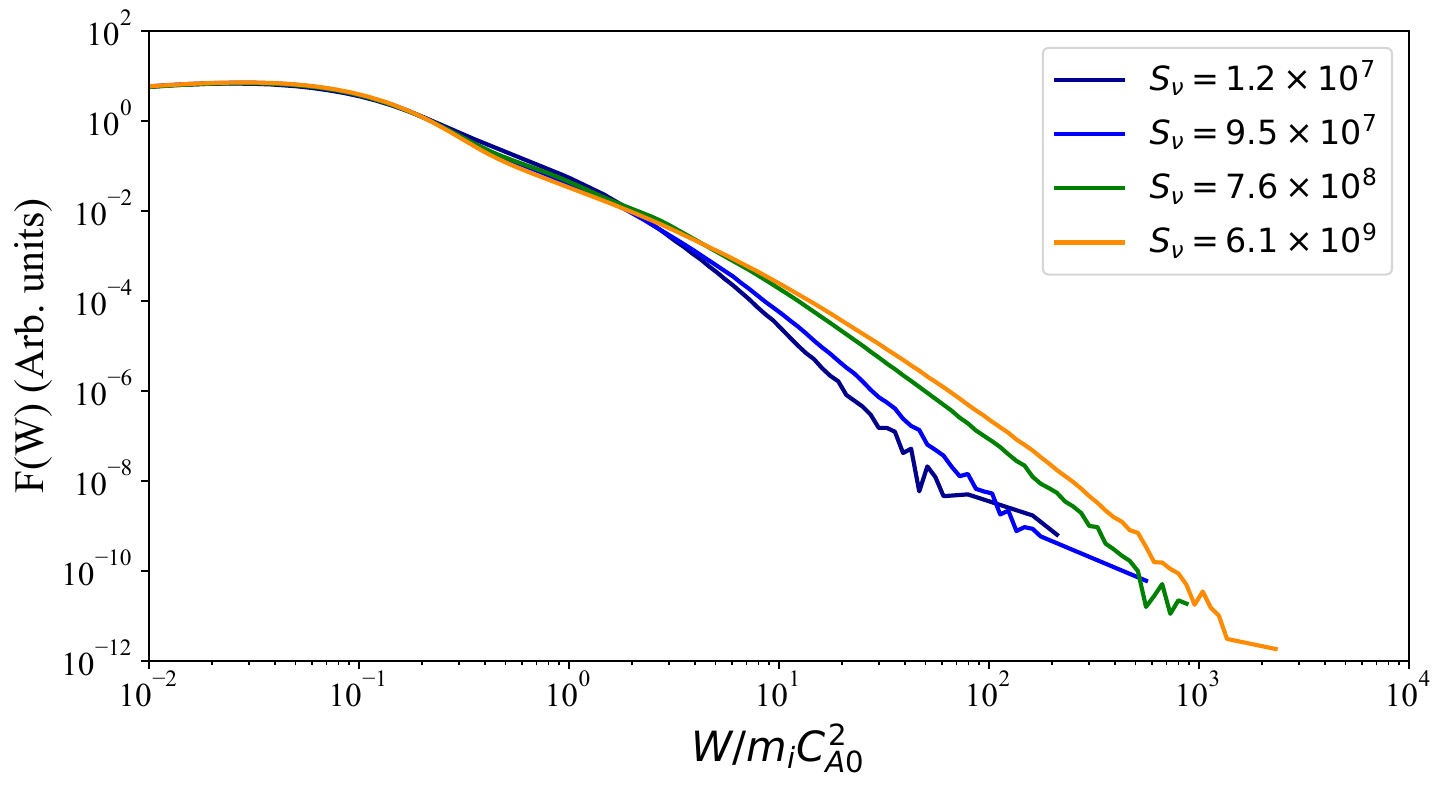}
\caption{The late-time distributions of energetic protons in a simulation with \( B_g / B_0 = 0.25 \) for various values of \(S_{\nu}\) (effective system size) reveal that the power-law indices of the spectra vary weakly with the domain size. The rightmost point of each curve marks the one-count level, indicating the flux when there is just one particle count per energy bin. The notable difference is that as the size of the system increases, the maximum energy of the protons also increases.
\label{fig:box_size}}
\end{figure}

In addition, we explore the late-time distributions of energetic protons for a guide field \( B_g / B_0 = 0.25 \) and various values of \(S_{\nu}\) (effective system size). The results in Fig. \ref{fig:box_size} show that the power-law indices of the spectra are only weakly dependent on the system size for large domains. As the size of the system increases, the spectra become slightly harder and the upper limit in the particle energy gain increases. This is because in larger domains more flux ropes form in the reconnecting current sheet and it is the merger of those flux ropes that efficiently drives particle energy gain. 
 
\begin{figure}[H]
\centering
\includegraphics[width=0.8\columnwidth]{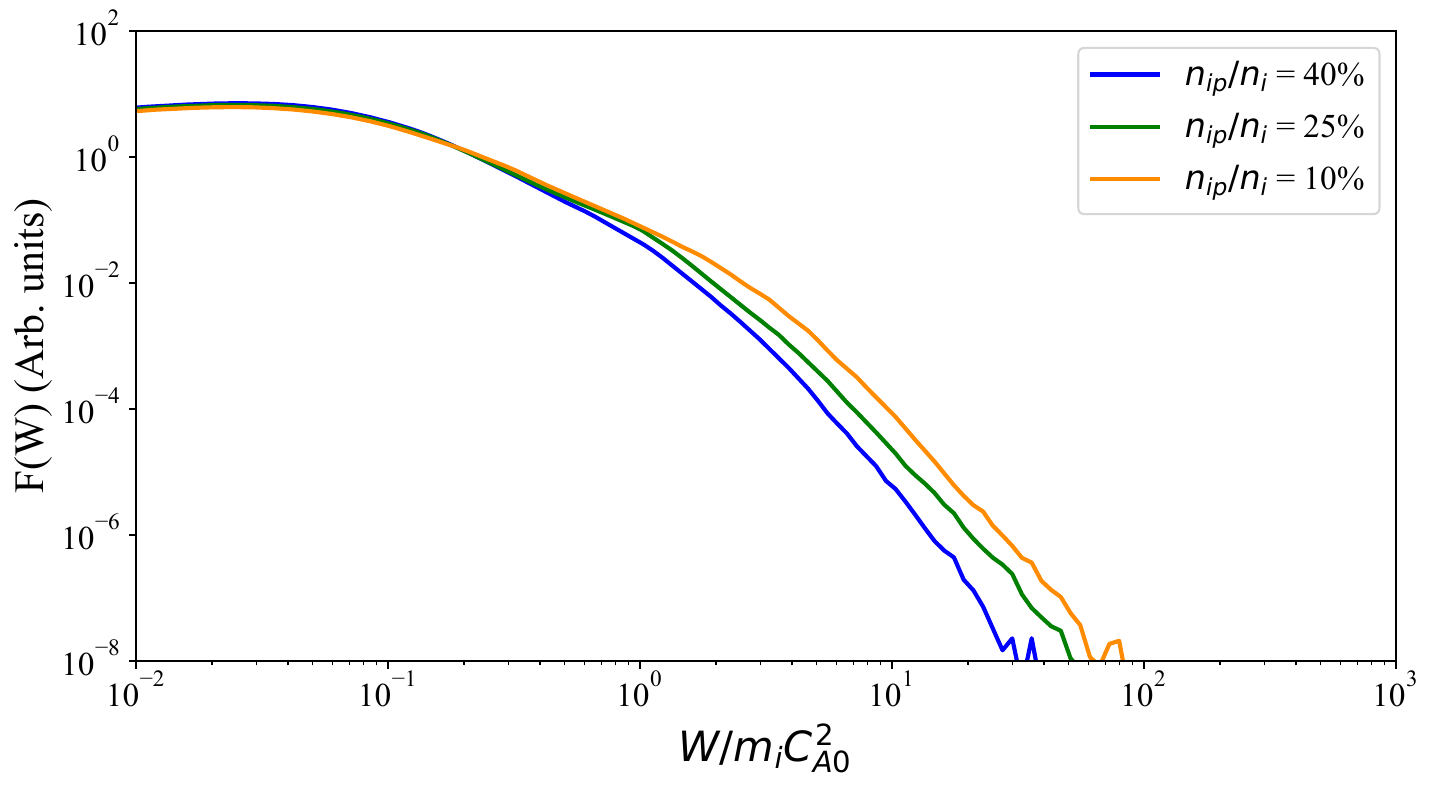}
\includegraphics[width=0.8\columnwidth]{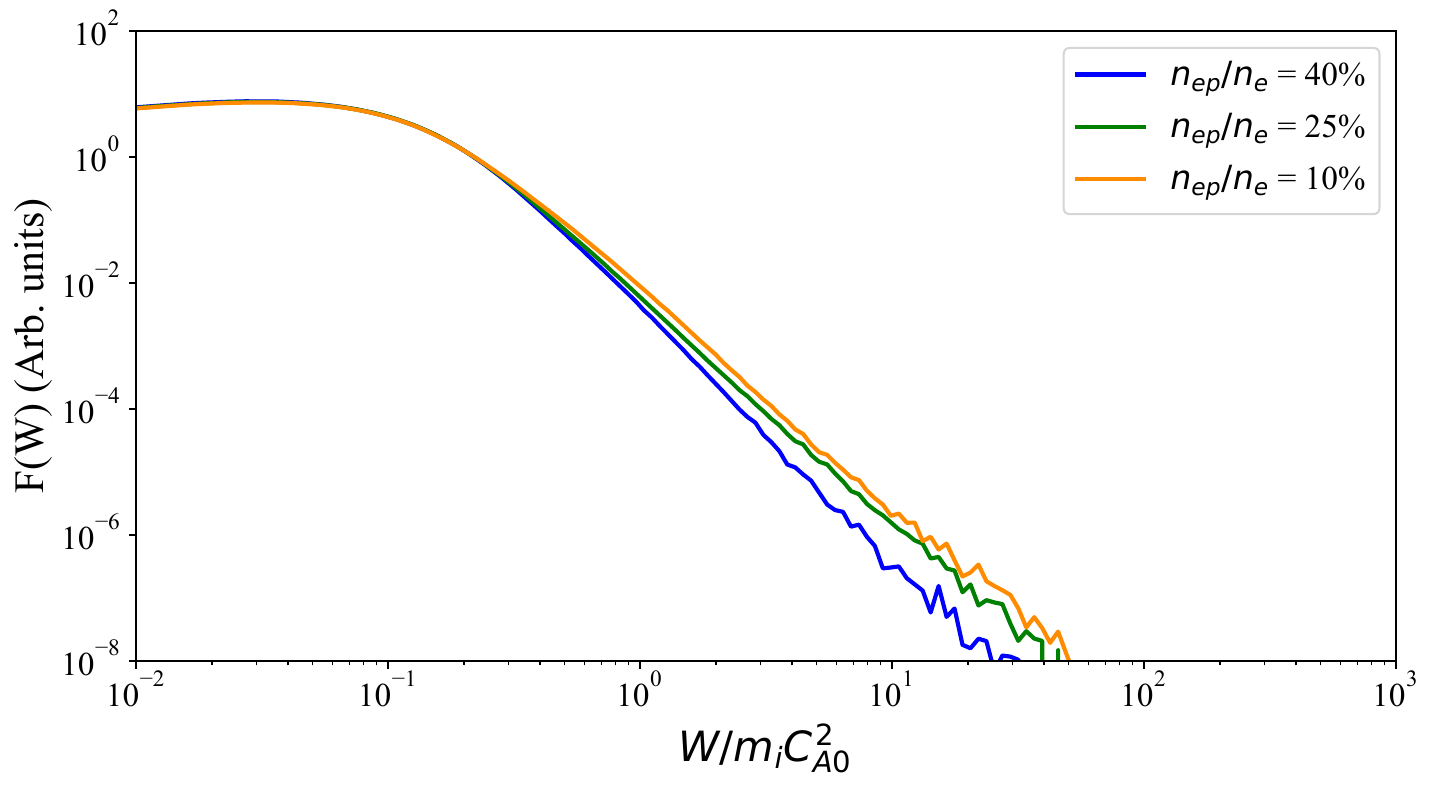}
\caption{The simultaneous distributions of protons (top) and electrons (bottom) from simulations with a guide field \( B_g / B_0 = 0.25 \) and different fractions of particle densities.  The power-law indices of the spectra for the different ratios are nearly the same, indicating that the simulation is insensitive to this ratio. The only notable difference is that as the fraction of particles decreases, the maximum achieved energy of each species increases modestly.
\label{fig:nfrac_ion}}
\end{figure}

A free parameter in the \textit{kglobal} model is the fraction of the particle electrons and protons compared with their respective fluid component \citep{Drake19,Arnold19,yin24}. An important question is whether reconnection simulations  are sensitive to these ratios. To address this question, we investigated the spectra of protons and electrons with a guide field \( B_g / B_0 = 0.25 \) and with different fractions of the particle densities. Figure \ref{fig:nfrac_ion} shows the late-time energy spectra of particle protons and electrons from three simulations with differing particle fractions. The power-law indices of the spectra are nearly identical, indicating that the simulation results are insensitive to this ratio. The only notable difference is that as the fraction of particles decreases, the maximum achieved energy of the particles increases modestly. This can be attributed to the fixed total energy in the simulation box. The particles gain more energy than their fluid counterparts with the parallel energy gain exceeding that of the perpendicular energy. Thus, when there are fewer particles in the simulation box, the particles can gain more energy before the firehose parameter acts to limit reconnection and energy gain.

\begin{figure}[H]
\centering
\includegraphics[width=\columnwidth]{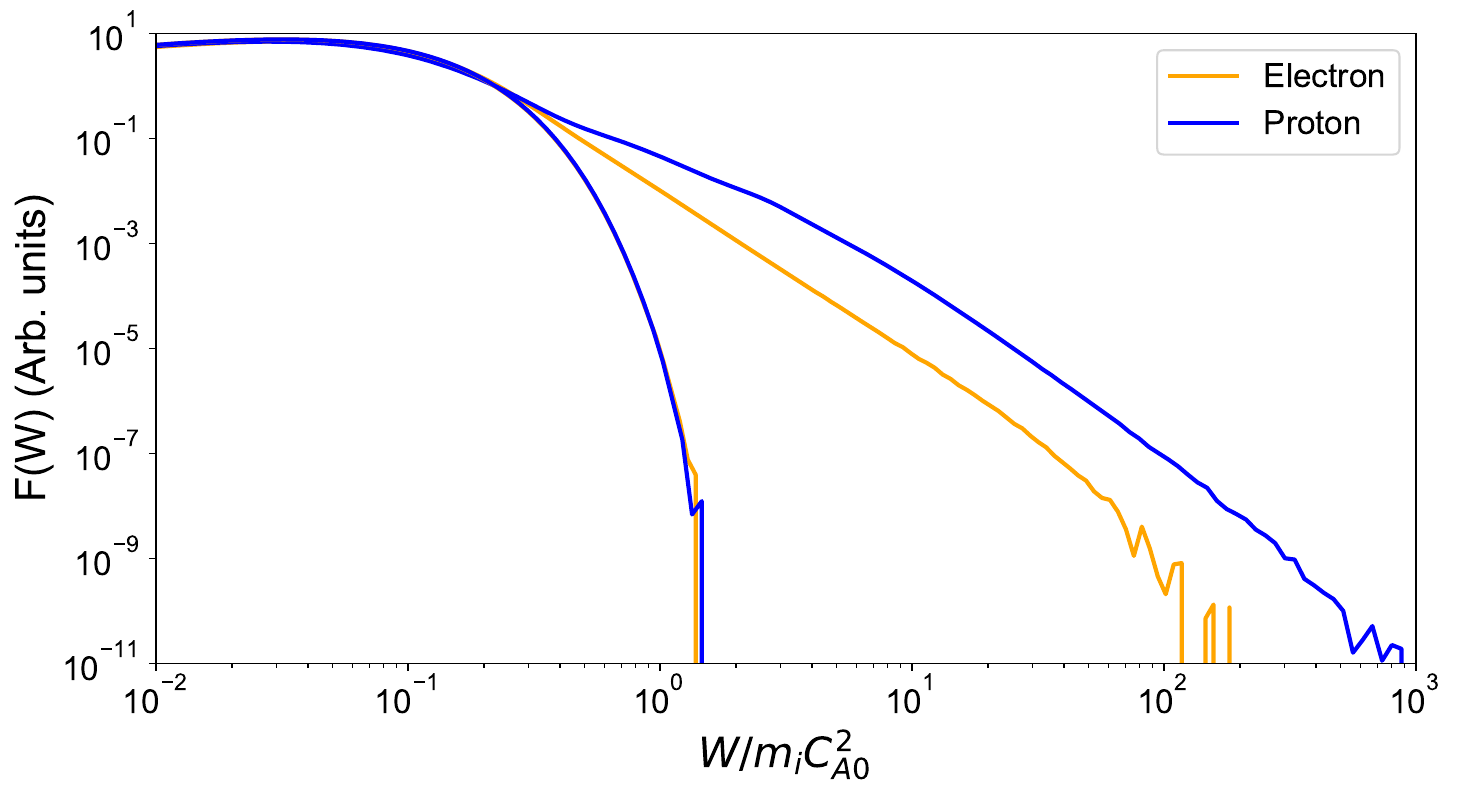}
\caption{The simultaneous distributions of protons (blue curves) and electrons (orange curves) for a guide field \( B_g / B_0 = 0.25 \). The curves on the left indicate the initial Maxwellian distributions for both particle types. The hard spectra are the late time results, showing extended power law distributions for both protons and electrons with powerlaw indices that are comparable. 
\label{fig:ei_spec}}
\end{figure}

We show in Figure \ref{fig:ei_spec} the initial and the late-time spectra of particle protons and electrons for the case \( B_g/B_0 = 0.25 \). The initial curves for both protons and electrons are Maxwellian distributions. Late-time spectra exhibit power-law tails for both particle types. The spectral indices of the powerlaws of the two species are nearly the same. Although the length of the power-law tail for the protons is shorter than that of the electrons, it extends nearly a full decade higher in energy. A key feature is the significantly greater energy gain of the protons compared to the electrons, which is consistent with the increased energy content of protons compared to electrons shown in Figure \ref{fig:fire}.

To compare our results to observations, consider that in solar flare current sheets the typical magnetic field magnitude is about 50 \(\mathrm{Gauss}\) and the density of protons is around \(10^{10}~\mathrm{cm}^{-3}\), which corresponds to an Alfv\'en speed of 1000~km/s. For our simulation, we initially set the temperatures of protons and electrons to be $0.0625 m_i C_{A0}^2$. Given that $m_i C_{A0}^2$ is approximately 10~keV in the solar case, this translates to an initial plasma temperature in the simulations of around $0.6$~keV (around 7 MK), which exceeds typical ambient coronal values in the absence of prior reconnection energy release. On the other hand, the late time energy spectra of both species extend to far larger energies, so the initial temperatures are not expected to significantly influence these late time spectra. The results shown in Figure \ref{fig:ei_spec} indicate that the maximum energy attained by energetic electrons can exceed $100m_iC_A^2 \sim 1$ MeV, while energetic protons can achieve energies in excess of $500m_iC_A^2\sim 5$ MeV.  These values are consistent with findings from gamma-ray flares documented by RHESSI \citep{Lin03}. A similar scaling of the  results from the largest simulation domain (see Figure \ref{fig:box_size}) suggests that the proton energy can exceed 20 MeV.

\begin{figure}[H]
\centering
\includegraphics[width=0.8\columnwidth]{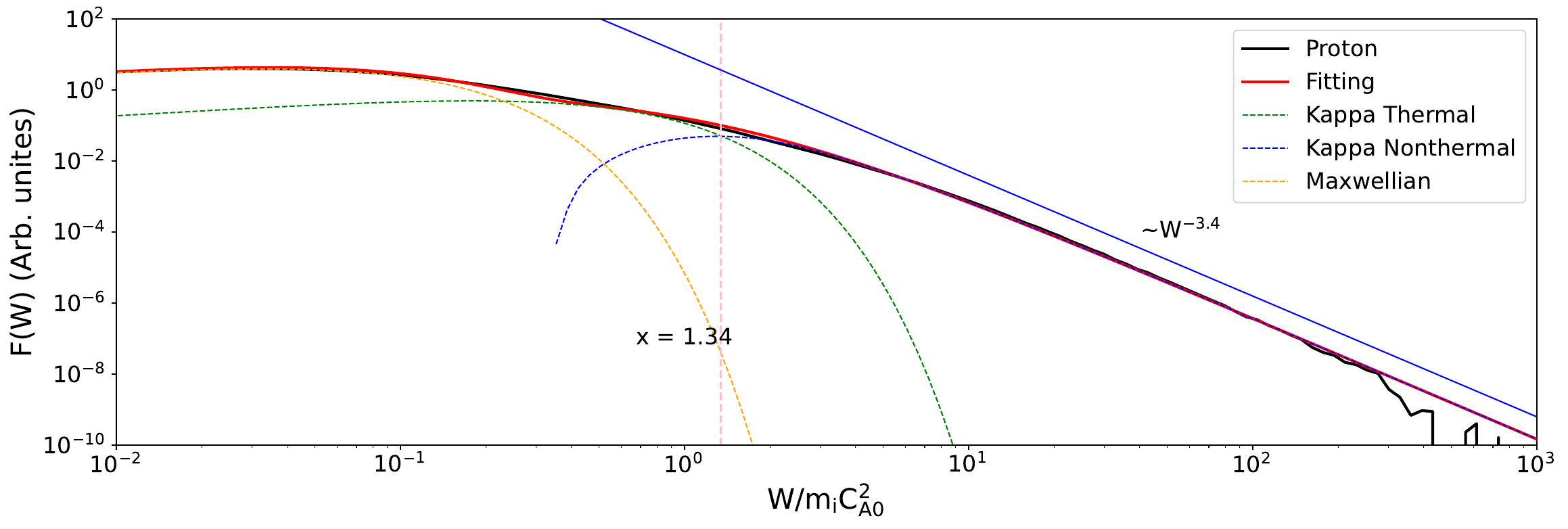}
\\
\includegraphics[width=0.8\columnwidth]{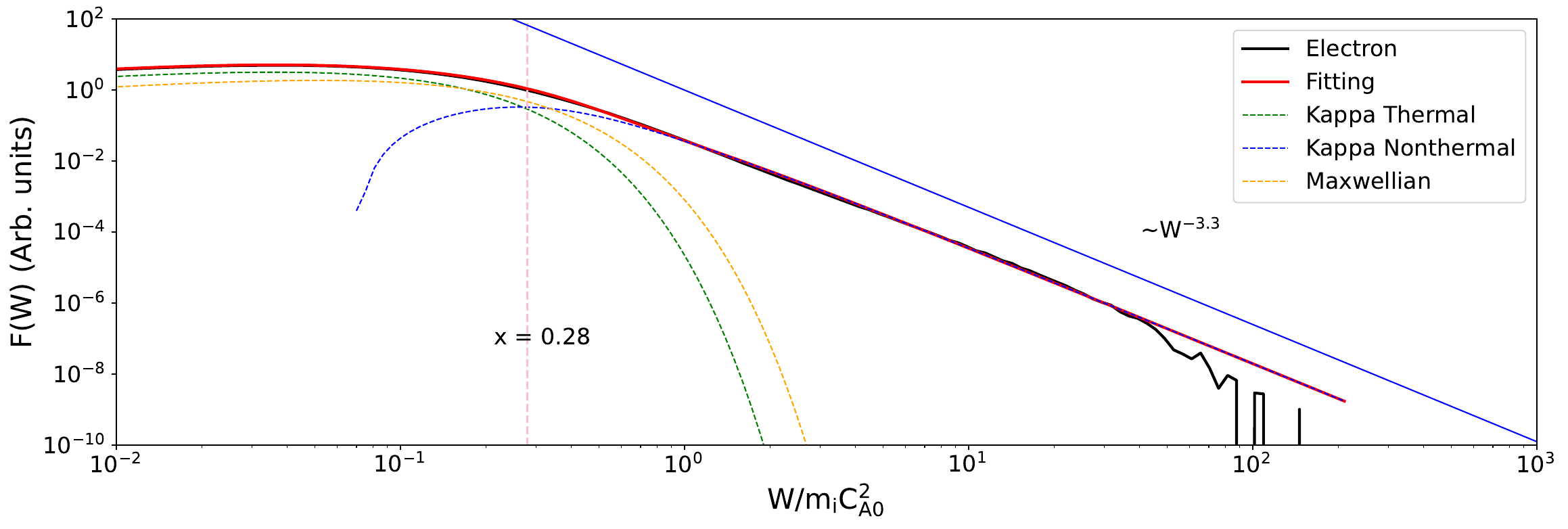}
\caption{Fitting and decomposition of the late time spectra of electrons and protons. The solid black curves are the proton (top) and electron (bottom) distribution functions for a guide field of 0.25. The solid red curves show the fit to a sum of a Maxwellian and a kappa distribution. The green dashed curve is the thermal part of the kappa distribution, while the blue dashed curve is the non-thermal part of the kappa distribution. The yellow curves are the Maxwellian distributions of the two species. The pink vertical line indicates the energy where the nonthermal and thermal components of the kappa distribution are equal. The solid blue line is the asymptote for the distribution function in the power law region.}
\label{fig:fit}
\end{figure}

We next explore the details of particle heating and energization by examining the relative numbers and energy content of the non-thermal and hot thermal particles. The spectra are modeled with the sum of a kappa \citep{vasyliunas68} and a Maxwellian distribution. The kappa distribution turns into a powerlaw at high energy which can represent the extended powerlaw tails seen in the simulation data \citep{Oka13}. This combination particularly useful in modeling the distributions of particles in space and astrophysical plasmas where non-thermal behaviors are observed \citep{kavsparova09,Krucker10,Oka13}.

\begin{figure}[H]
\centering
\includegraphics[width=0.8\columnwidth]{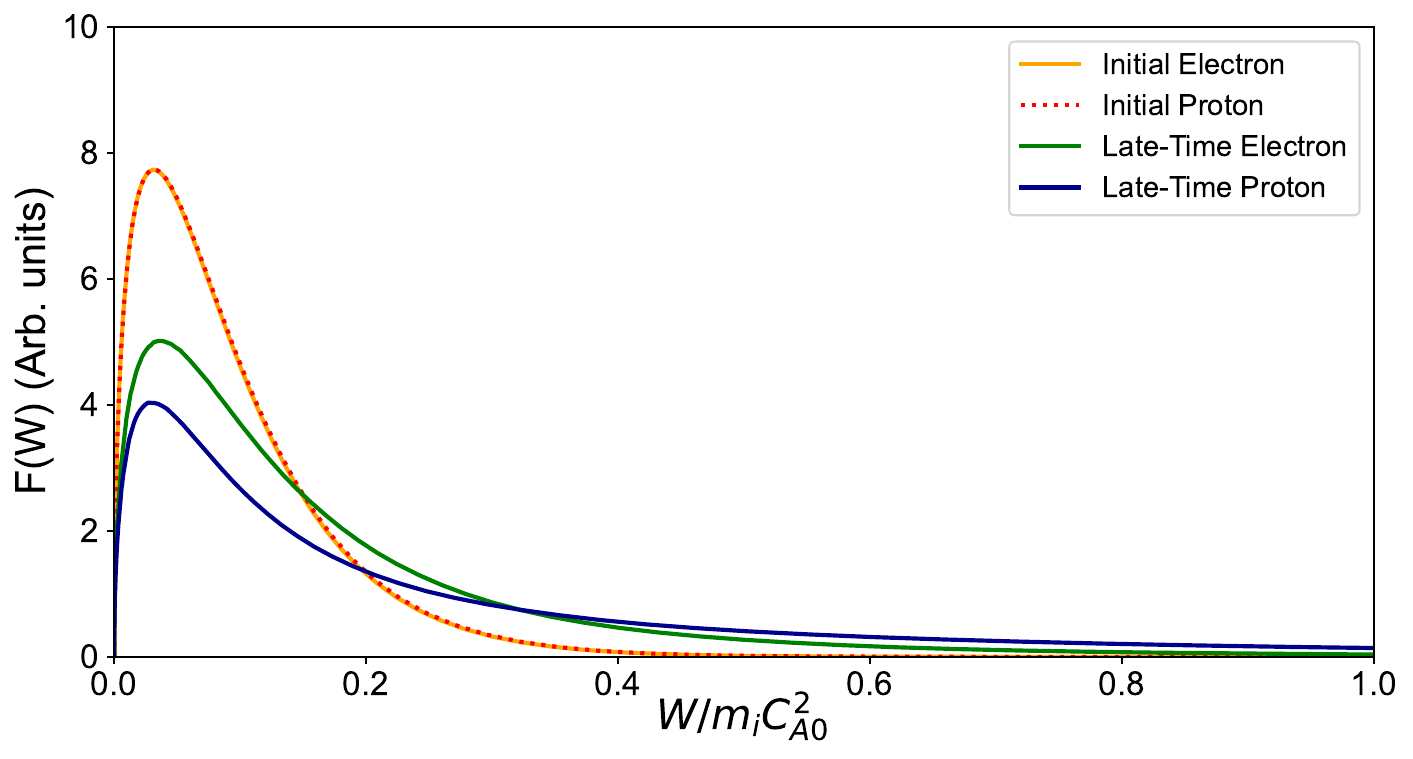}
\caption{The comparison between the initial spectra and late-time spectra of protons and electrons is presented on a linear-linear scale to more clearly reveal the changes in the particle distributions at low energy.}
\label{fig:init_late_comparison}
\end{figure}

In Figure \ref{fig:fit} we show \( F(W) \) (black) for protons and electrons at late time after the distribution functions have reached a steady state. The data is taken from within the large late-time island shown in Figure \ref{fig:evo}(e). Thus, the spectra shown include only particles that have undergone heating and energization during reconnection. Shown in Figure \ref{fig:init_late_comparison} are the intial and final electron and proton distributions from the same large island but using a linear scale. This figure more clearly shows the changes in the distribution of low energy particles as they are heated during reconnection. The spectra in Figure \ref{fig:fit} are fit (solid red) by the sum of a Maxwellian distribution and a kappa distribution \citep{Arnold21}. This procedure is designed to capture the characteristics of both hot thermal and nonthermal particle populations. From this fitting approach, we extract several parameters:

\begin{enumerate}

    \item The spectral index $\delta^{\prime}$ of the powerlaw of the non-thermal particles.
    \item The relative densities densities and total energy of the hot thermal and non-thermal electrons and protons. 

    \item The break point energy (vertical line in Figure \ref{fig:fit}) , expressed in  units of \(m_iC_{A0}^2\), that marks the transition from thermal to non-thermal particle dominance in the spectra. At the break point the number densities of the hot thermal and non-thermal components of the kappa distribution are equal (See Appendix \ref{sec:kappa} for a detailed explanation of the procedure used to separate the hot thermal and non-thermal components of the kappa distribution).  To the left of the break point, the thermal component dominates, while to the right, the thermal components decrease sharply, and the non-thermal component dominates.
 \end{enumerate}



\begin{deluxetable*}{cccccc}
\tablewidth{0pt}
\tablecaption{Density Distributions for Particle Acceleration at Late Time \label{tab:density}}
\tablehead{
  \colhead{Type} & 
  \colhead{Guide Field} &
  \colhead{Maxwellian} & 
  \colhead{Kappa Total} & 
  \colhead{Kappa Maxwellian} & 
  \colhead{Kappa Nonthermal} \\[-1.5ex]
  \colhead{} & 
  \colhead{(in $B_0$)} &
  \colhead{(\%)} & 
  \colhead{(\%)} & 
  \colhead{(\%)} & 
  \colhead{(\%)}
}
\startdata
Electron & 0.25 & 38.2 & 61.8 & 45.1 & 16.7 \\
Electron & 0.50 & 68.0 & 32.0 & 26.8 & 5.2 \\
Electron & 0.75 & 84.3 & 13.7 & 12.4 & 1.3 \\
Proton & 0.25 & 51.7 & 48.3 & 35.8 & 12.5 \\
Proton & 0.50 & 43.1 & 56.9 & 46.7 & 10.2 \\
Proton & 0.75 & 54.5 & 45.5 & 40.3 & 5.2 \\
\enddata
\end{deluxetable*}


\begin{deluxetable*}{ccccccccc}
\tablewidth{0pt}
\tablecaption{Relative Energy and Spectra Parameters for Particle Acceleration at Late Time\label{tab:combined}}
\tablehead{
  \colhead{Type} & 
  \colhead{Guide} & 
  \colhead{Total} & 
  \colhead{Maxwellian} &
  \colhead{Kappa} & 
  \colhead{Kappa} & 
  \colhead{Thermal} & 
  \colhead{Nonthermal} & 
  \colhead{\makebox[1.2cm]{Powerlaw}} \\[-2.0ex]
  \colhead{} & 
  \colhead{Field} & 
  \colhead{} & 
  \colhead{} & 
  \colhead{Maxwellian} & 
  \colhead{Nonthermal} &
  \colhead{Temperature} &
  \colhead{Break Point} &
  \colhead{Index} \\[-1.5ex]
  \colhead{} & 
  \colhead{(in $B_0$)} & 
  \colhead{(in $m_i C_{A0}^2$)} & 
  \colhead{(in $m_i C_{A0}^2$)} & 
  \colhead{(in $m_i C_{A0}^2$)} & 
  \colhead{(in $m_i C_{A0}^2$)} &
  \colhead{(in $m_i C_{A0}^2$)} & 
  \colhead{(in $m_i C_{A0}^2$)} &
  \colhead{($\delta^{\prime}$)} 
}
\startdata
Electron & 0.25 & 0.202 & 0.059 & 0.048 & 0.095 & 0.086 & 0.28 & 3.3\\ 
Electron & 0.50 & 0.158 & 0.101 & 0.033 & 0.024 & 0.094 & 0.35 & 5.4\\
Electron & 0.75 & 0.135 & 0.113 & 0.016 & 0.006 & 0.087 & 0.50 & 8.8\\
Proton & 0.25 & 0.581 & 0.050 & 0.191 & 0.340 & 0.184 & 1.34 & 3.4\\
Proton & 0.50 & 0.295 & 0.035 & 0.141 & 0.119 & 0.132 & 0.72 & 4.9\\
Proton & 0.75 & 0.200 & 0.051 & 0.104 & 0.045 & 0.109 & 0.86 & 7.7\\
\enddata
\end{deluxetable*}

Table \ref{tab:density} documents how the strength of the guide field affects the proportion of particles that comprise the Maxwellian and kappa distributions, with the latter further subdivided into thermal and non-thermal components.  As the guide field increases, the proportion of non-thermal particles, indicated by the kappa non-thermal percentage, consistently decreases. For electrons, the non-thermal component reduces significantly from a higher percentage at a lower guide field to a much smaller fraction at the higher guide field. A similar trend is observed for protons, where the non-thermal proportion diminishes as the guide field strength increases. This trend establishes that a strong guide fields suppresses the generation of non-thermal electrons and protons, which is consistent with the findings for energetic electrons in \cite{Arnold21}.

Table \ref{tab:combined} shows the relative energy of thermal and non-thermal electrons and protons at late time, again as a function of guide field strength. In this data the total number density of particles is normalized to unity and the numbers in the table reflect the energy content of the various components in units of $m_iC_A^2$. As the guide field increases, the energy in the non-thermal component of the kappa distribution decreases for both electrons and protons. For electrons, the kappa non-thermal energy decreases from 0.095 at a guide field of 0.25 to 0.006 at a guide field of 0.75. For protons, the kappa non-thermal energy decreases from 0.340 at a guide field of 0.25 to 0.045 at a guide field of 0.75. The downstream thermal temperatures are also shown. For electrons the thermal temperature, represented as a weighted sum of the Maxwellian and Kappa Maxwellian distributions, remains comparable for different guide fields. The temperature increase is modest compared with the initial value of $0.0625m_iC_{A0}^2$. The heating is driven by the macroscale parallel electric field as explored by \citet{Arnold21}. In contrast, protons exhibit a much larger increase in temperature as has been documented in the Earth space environment 
\citep{Phan13a,Phan14,Oieroset23}. For the lowest guide field in Table \ref{tab:combined} the ratio of the temperature increment of the protons to electrons is around five, which is slightly smaller than in measurements in the Earth space environment \citep{Phan14}. Also noticeable is the decrease in thermal temperature of protons with increasing guide field. We emphasize that because the data in Table \ref{tab:combined} is normalized to $m_iC_{A0}^2$, the temperature increments of both species scales with this parameter as in the observations \citep{Drake09,Phan13a,Phan14,Oieroset23}. 

Also shown in Table \ref{tab:combined} is the breakpoint energy, which marks the transition from thermal to non-thermal particle dominance in the spectra. The depedence of the breakpoint energy on the guide field behaves differently for electrons and protons. For electrons, as the guide field increases, the break point shifts to higher values, while for protons the break point first decreases with increasing guide field and then the trend reverses. The powerlaw index also increases with the guide field, reflecting a steep drop-off in the number of non-thermal particles at higher energies. 



\section{Conclusion} \label{sec:conclusion}
In this paper, we present the results of  numerical simulations of the self-consistent energization of electrons and protons during magnetic reconnection in a macroscale system. We calculate the time evolution of the macroscale system, starting from a thin current sheet, progressing through the formation of multiple flux ropes and their merger, culminating in the formation of a large merged island at late time. Electrons and protons gain energy through Fermi reflection as multiple flux ropes grow and merge. At late time the energetic particles have mostly accumulated into the large merged island as seen in earlier electron-only simulations \citep{Arnold21}. We also present an example of the firehose stability parameter at late time. The core of magnetic flux ropes approach firehose marginal stability so that the magnetic tension within flux ropes (the drive mechanism for reconnection) is suppressed. It is through the firehose parameter that energetic protons and electrons feed back onto the MHD fluid during magnetic reconnection. 

The proton and electron spectra both form power-law distributions at high energy with comparable spectral indices. As the magnitude of the guide field decreases, both protons and electrons exhibit a harder energy spectra so that there are larger numbers of high-energy particles. Similar trends have been previously reported \citep{Dahlin14,Arnold21,zhang21}. The power-law slopes are generally insensitive to the system size. However, the highest energy of non-thermal particles increases in larger systems, which reflects the larger number of flux ropes undergoing merger to drive particle energy gain. The powerlaw spectra of both species extend nearly three decades in energy for large domains. Our results also indicate a distinct difference in energy gain between protons and electrons: protons gain more energy and their maximum energy exceeds that of electrons.

Energy is normalized to $m_iC_{A}^2$ in our simulations.  A typical value of this parameter in a solar flare is 10 keV (see Sec.~\ref{sec:results}). Thus, Table \ref{tab:combined} suggests that hot thermal electrons and protons in flares should temperatures of the order of 0.3$keV$ (around 0.4 MK) and 1.2$keV$ (1.4MK), respectively. Further, our simulations suggest that energetic electrons can reach energies as high as $10^2m_iC_{A}^2\sim 1 $MeV, while energetic protons can reach beyond $10^3m_iC_{A}^2\sim 10$ MeV. The simulations further reveal that the break point, which is the transition from thermal to non-thermal particle dominance in the particle spectra, is around  $1.34m_iC_{A}^2\sim 13 $keV for protons and $0.28m_iC_{A}^2\sim 3$ keV for electrons. The powerlaw spectra of protons should extend down to far lower energies than the typical 1 MeV minimum proton energy that is measurable in flares \citep{Lin03,Emslie12}. Of course, the powerlaw spectra of a variety of ion species from impulsive flares have been documented in spacecraft measurements in the solar wind \citep{Reames99,Mason07}. This suggests that the total energy content of protons in solar flares is likely to be greater than previously expected \citep{Emslie12}. 

Magnetic reconnection in low density events in the Earth's magnetotail are associated with high values of $m_iC_A^2$ with Alfv\'en speeds reaching values above 1000km/s. Powerlaw distributions of electrons \citep{Oieroset02,Ergun20} as well as protons \citep{Ergun20} in such events have been documented. {\it In situ} measurements of reconnection outflows have documented that electron \citep{Phan13} and proton heating \citep{Drake09,Phan14} scale with $m_iC_A^2$ and recently the observations have been found to extend to the energetic tail events, with electron temperature increments reaching several keV \citep{Oieroset23}. Simulations carried out with particle-in-cell codes also support this scaling of particle energy gain during reconnection \citep{Haggerty15}. Since the energy normalization parameter in our simulations is $m_iC_A^2$, the results of our simulations shown in Table \ref{tab:combined} also support this scaling. The {\it in situ} observations have established that proton bulk heating during reconnection is substantially greater than that of electrons by a factor of around 8. The data in Table \ref{tab:combined} indicates that the ratio of the increments of proton to electron temperature is around 5 for the smallest guide field simulation (the observations are mostly limited to weak guide field events), somewhat less than in the observations. On the other hand, in the Earth space environment the protons typically have initial temperatures well above that of electrons. This might account for their greater energy gain since the Fermi heating and acceleration is strongest for particles with greater energy. 

Thus, simulations with the \textit{kglobal} model can be carried out to explore reconnection-driven heating and energization in macroscale systems. By using measured values of $m_iC_{A}^2$, the initial temperatures of protons and electrons, and ambient guide fields, the simulation model can produce results for direct comparison with observational data from solar flares and {\it in situ} measurements from the Earth space environment and and the solar wind. 

A caveat to the present simulation results are that they have been carried out in 2D systems. It is well known that reconnection-driven particle energization is stronger in 3D than in 2D. In 3D systems energetic particles can escape from isolated flux ropes where particle energization has ceased to sample active reconnection sites \citep{Dahlin15}. Further, magnetotail events are known to be strongly turbulent \citep{Ergun20}. How this turbulence impacts energization remains an open question that requires further exploration. 

\begin{acknowledgments}
We acknowledge support from the FIELDS team of the Parker
Solar Probe (NASA Contract No. NNN06AA01C), the NASA Drive
Science Center on Solar Flare Energy Release (SolFER) under Grant No. 80NSSC20K0627, NASA Grant Nos. 80NSSC20K1277 and
80NSSC22K0352, and NSF Grant No. PHY2109083. The
simulations were carried out at the National Energy Research
Scientific Computing Center (NERSC). The data used to perform
the analysis and construct the figures for this paper are preserved at the NERSC High Performance Storage System and are available upon request.
\end{acknowledgments}

\appendix

\section{Spectral Fitting}
\label{sec:kappa}
The \textit{kappa distribution} is frequently used to characterize velocity distributions that deviate from the traditional Maxwellian due to the presence of a pronounced high-energy tail. This characteristic is indicative of plasmas that harbor a significant population of non-thermal particles, a phenomenon commonly observed in environments such as the solar corona and other astrophysical settings.  It converges to a Maxwellian distribution as the kappa parameter (\(\kappa\)) becomes large, while for small \(\kappa\) the non-thermal population, as manifested by the presence of an elongated high-energy tail, increases.
The mathematical expression for the kappa distribution function for particle energy (\(W\)) is:

\begin{equation}
F_\kappa (W) = A \left(1 + \frac{W}{T_\kappa (\kappa - 3/2)}\right)^{-(\kappa + 1)}\sqrt{W}
\label{Fkappa}
\end{equation}
where \(A\) is a normalization constant, \(W\) is the particle energy, \(\kappa\) is a shape parameter that influences the fraction of particle in the non-thermal tail and $3T_\kappa /2$ is the total integrated energy in the $\kappa$ distribution. \cite{Oka13} demonstrated how kappa distributions can be decomposed into thermal and non-thermal components \citep[see also][]{oka15a}.
\begin{itemize}
    \item \textbf{Thermal Component:} This component is characterized by velocities near the peak of the distribution, where it resembles a Maxwellian. The temperature of the Maxwellian component of the kappa distribution is:       
    \begin{equation}
    T_{\kappa M} = T_\kappa (\kappa -3/2)/\kappa, 
    \end{equation}
    so that 
    \begin{equation}
       F_{\kappa M} (W) = A' e^{-W/T_{\kappa M}}\sqrt{W}, 
    \end{equation}
    where the normalization $A'$ is shifted slightly from that in Equation \ref{Fkappa} as discussed in \cite{Oka13}. 
    \item \textbf{Non-Thermal Component:} This component includes a high-velocity tail, where the distribution decays more slowly with increasing energy than a Maxwellian distribution. It describes particles with energies above the average thermal energy, with the prominence of this non-thermal tail being regulated by the value of \(\kappa\). Lower \(\kappa\) values denote a larger population of high-energy particles. The non-thermal spectrum is defined by
    \begin{equation}
        F_{\kappa nt}=F_\kappa-F_{\kappa M}
    \end{equation} 
The number of non-thermal versus thermal particles is given by \citep{Oka13},  
    \begin{equation}
N_{\kappa nt} = N_\kappa - N_{\kappa M} =
N_\kappa \left( 1 - 2.718 \frac{\Gamma(\kappa + 1)}{\Gamma(\kappa - 1/2)} \kappa^{-3/2} \left( 1 + \frac{1}{\kappa} \right)^{-(\kappa + 1)} \right)
\end{equation}
The energy content of the non-thermal particles is 
\begin{equation}
<W_{\kappa nt}> = <W_\kappa> - <W_{\kappa M}> = \frac{3}{2} T_\kappa - \frac{3}{2} T_{\kappa M}
\end{equation}
\end{itemize}
More details about the kappa distribution fitting can also be found in \cite{Arnold21}.

An important additional characteristic of the particle spectra is the low energy boundary (the "break point") of the powerlaw distribution $W_{nt}$. We take this to be the energy at which $F_{\kappa nt}(W_{bp})=F_{\kappa M}(W_{bp})$. The values of $W_{bp}$ for electrons and protons are given in Table \ref{tab:combined} for several values of the guide field. 

\bibliography{reference}{}
\bibliographystyle{aasjournal}
\end{CJK*}
\end{document}